\documentstyle[11pt,newpasp,twoside,epsf]{article}
\def\edcomment#1{\iffalse\marginpar{\raggedright\sl#1\/}\else\relax\fi}
\reversemarginpar

\begin{document}
\title{Perturbations to Stellar Structure in 2D: Stellar Rotation and Heating in X-ray Binaries.}
\author{John J. Eldridge, Pascale Garaud and Christopher A. Tout}
\affil{Institute of Astronomy, Madingley Road, Cambridge, CB3 0HA, England.}

\begin{abstract}
We have developed a numerical code with which we study the effects of 2D perturbations on stellar structure. We present new numerical and analytical results on the heating of a main-sequence star in a binary system by its companion.

\end{abstract}

\section{Introduction.}

Numerical study of stellar structure and evolution is now a well established subject with a long history. Today there is a plethora of 1D stellar evolution codes, from which astrophysicists can choose. All these models rely on the assumption that the star is spherically symmetric.

This assumption has served well for many years. However it is insufficient in cases that require stars to be non-symmetric. Stellar rotation, accretion and binary systems all add asymmetry to a spherical star. The required computational power is now available to look at these perturbations.

We have been working to develop a code that calculates 2D stellar structures in a non-computationally demanding way. So far we have studied perturbations to structure only but we plan to move into 2D stellar evolution. The two areas we have studied are rotation of solar-like stars and heating of a star by a companion star. The results are presented in this paper.

\section{Numerical Method.}

The method we use was developed by Garaud (2002). A 1D model in hydrostatic equilibrium is taken from a standard 1D evolution code and used as the underlying state for perturbation to the structure in 2D. The perturbative equations are linear with respect to thermodynamic quantities, with full non-linearity kept for the flow perturbations. The equations are

\begin{equation}
\rho {\textbf{u}} \cdot \nabla {\textbf{u}} = - \nabla \tilde{p} - \rho \nabla \tilde{\Phi} - \tilde{\rho} \nabla \Phi + \nu \nabla^{2} {\textbf{u}} + \frac{1}{3} \nu \nabla (\nabla \cdot {\textbf{u}}),
\end{equation}

\begin{equation}
\nabla^{2} \tilde{\Phi} = 4 \pi G \tilde{\rho},
\end{equation}

\begin{equation}
\frac{ \tilde{p} }{p} = \frac{ \tilde{\rho} }{\rho} + \frac{ \tilde{T} }{T},
\end{equation}

\begin{equation}
\nabla \cdot (\rho \textbf{u}) = 0,
\end{equation}
and
\begin{equation}
\rho T \textbf{u} \cdot \nabla s = k \nabla^{2} \tilde{T},
\end{equation}
where \textbf{u} is fluid velocity, $p$ the pressure, $\rho$ the density, $\Phi$ the gravitational potential, $\nu$ the kinematic viscosity, $T$ the temperature, $k$ the thermal conductivity and $s$ the entropy. The variables with a tilde are the perturbed variables and those without are taken from the 1D hydrostatic equilibrium model. We look for solutions to these equations that are stationary and axisymmetric. Note that we do not perturb the energy generation or opacities. The solution method is to use finite differences in the radial direction and expansion upon Chebyshev polynomials (essentially a Fourier decomposition) in the latitudinal direction.

Once this system is setup all that remains is to impose the boundary conditions appropriate to the system considered. 

\section{Stellar Rotation.}

Stellar Rotation is an old problem (Eddington 1925; Sweet 1950). Commonly used models examine meridional currents in stars that are rotating slowly, that is more slowly than the Sun. There are two main effects of rotation on stellar evolution, the increase or decrease of the amount of contraction or expansion of a star during its evolution and the induction of extra mixing of material within the star, especially within regions that are convectively stable. This has been recently reviewed by Pinsonneault (1997).

Garaud (2002) studied the meridional currents in the radiative zone of solar-like stars set up by the differential rotation of an outer convective zone. To investigate the problem a radiative zone model was taken from a 1D standard solar model (Christensen-Dalsgaard et. al. 1991) and perturbed by forcing rotation on the outer boundary and the equations were solved within the radiative zone of the star, with the assumption that the convection is not affected by the radiative zone because the driving forces are larger than any effect from the radiative-zone flow.

The boundary condition imposed differential rotation at the surface of the radiative zone of the form

\begin{equation}
\Omega_{\textrm{surface}} (\theta) = \sum_{n} \Omega_{n} \cos^{n} \theta
\end{equation}
along with stress continuity across the surface and $\nabla^{2} T = 0$ outside. For further details see Garaud (2002).

\section{Heating of Main-Sequence Stars in X-ray Binaries.}

It is possible to study different problems with the same method. Essentially only the boundary conditions require alteration. Therefore we adapted the code to study heating of a main-sequence star in a binary system by X-rays emitted by an accreting neutron-star companion. Many such X-ray binaries are known to exist.

Observational evidence that something non-1D is occurring comes from HZ~Herculis where there are variations in the light curve, especially at the minimum when the X-ray source is eclipsed, that cannot be explained by a main-sequence star with a homogeneous surface temperature. It is generally assumed that the external radiation that falls on the star's surface is thermalized and re-emitted, with some of the energy emitted from the unilluminated side. To achieve this there must be some form of circulation current in the star's envelope.

There have been previous attempts to model this situation, Kippenhahn \& Thomas (1978) assumed that the flux was thermalized very near to the surface and found that the induced flow penetrated to a depth of 20\% of the star's radius, with a 10\% increase in the illuminated hemisphere temperature and 2\% of the extra flux emitted from the unilluminated surface. The velocities of the vertical flows are subsonic, while the flows parallel to the surface are likely to be supersonic.

Later, K\i rb\i y\i k (1982) produced a similar model but with a much shallower penetration of only a few percent of the star's radius. These analyses of the reflection effect have been used to explain many light curves in close binary systems, with the solution of Kippenhahn \& Thomas (1978) being the most commonly citied.

\subsection{Boundary Conditions}

Careful thought must be given to formulation of the boundary conditions because there are many possibilities for boundary conditions that add extra energy to the surface of the star. The X-ray flux is absorbed in a thin layer near the surface. The column density required is of the order of $0.1 \to 1 \, \rm{g \, cm^{-2}}$. This is reached at a very shallow depth in the stellar surface, under the assumption that there is no absorption in the stellar atmosphere.

To a first approximation we can say that, at the surface, we are adding extra energy with a boundary condition

\begin{equation}
-k \frac{\partial T}{\partial r} = \sigma_{\rm{B}} T^{4} - \alpha f(\theta) \cos \theta.
\end{equation}
The function $f$ is defined as
\begin{equation}
f(\theta) =  \left\{ 
\begin{array}{ll}
1 & -\frac{\pi}{2} < \theta < \frac{\pi}{2},\\
0 & \textrm{otherwise},\\
\end{array} \right.
\end{equation}
where $\theta$ is the latitude, $\alpha$ the strength of the X-ray flux, $\alpha = L_{X} / (4 \pi d^{2})$. Because we are dealing with small perturbations we expand $T \to T + \tilde{T}$ and assuming $\tilde{T}$ is small we obtain
\begin{equation}
-k \frac{\partial \tilde{T}}{\partial r} = 4\sigma_{\rm{B}} T^{3}\tilde{T} - \alpha f(\theta) \cos \theta.
\end{equation}
This boundary condition can also be used if the illuminating star is a main-sequence star.

We also have conditions on the flow such that
\begin{equation}
u_{r}(r=1)=0
\,\, {\rm{and}} \,\,
u_{\theta}(\mu =\pm1)=0,
\end{equation}
along with a stress free boundary on the flow at the surface. We have defined $\mu=\cos \theta$, r the dimensionless radial coordinate, $r=0$ is at the centre and $r=1$ is at the surface.

\subsection{Numerical Results}

\begin{figure}[!h]
\plotone{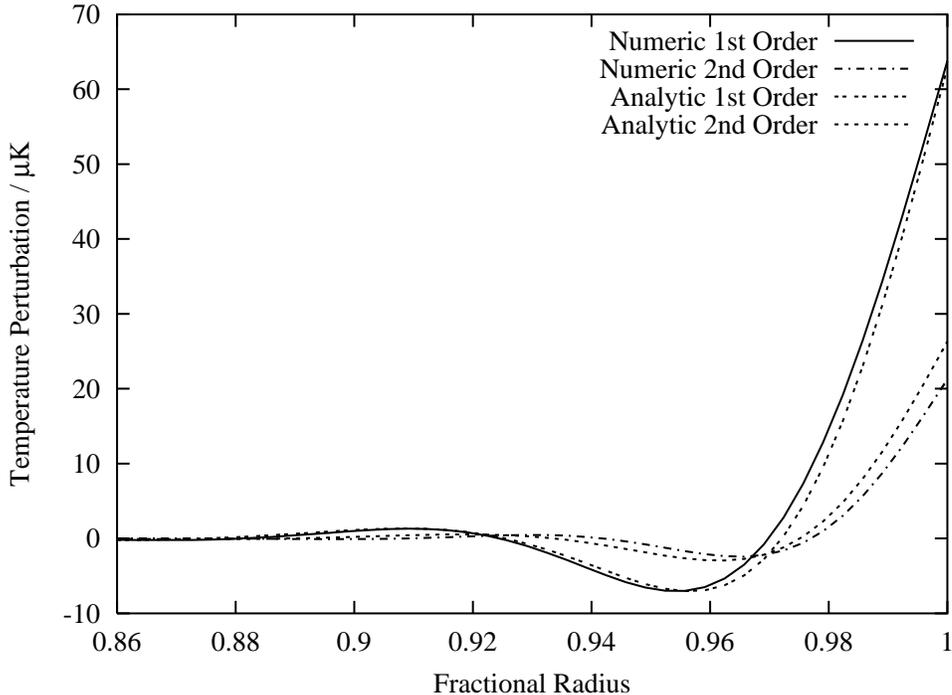}
\caption{Plot of the radial dependence of the temperature perturbations for the solar radiative zone (surface taken to be at $0.6 R_{\sun}$), the illuminating X-ray flux is $10^{35} \rm{erg \, s^{-1}}$ at 1 au. Notice the shallow penetration and the good agreement of analytic and numerical results.}
\end{figure}

Only one model, the solar radiative zone, has been studied successfully numerically. By this we mean that we have removed the upper part of the Sun so we are directly heating the region at $0.6 R_{\sun}$. This was done to test the numerics on a well-known background state and helped us develop analytical tools that can be generalised to any stellar radiative zone. The viscosity has also been boosted to resolve the thin thermo-viscous boundary layers.

We find that circulation is induced but limited to a thin boundary layer near the surface. Similarly to previous authors we find that there are two circulation cells. Figure 1 shows the temperature fluctuations versus depth. The perturbations only penetrate to a shallow depth. Figure 2 plots the value of $\psi$ which represents the nature of the induced flows. Radial velocities are subsonic but latitudinal velocities do reach supersonic values for illumination equivalent to a source of $10^{35} \rm{erg} \, \rm{s}^{-1}$ at $1$ au. The flow velocities are linearly dependent on the illuminating flux.

Because the effect of the heating is confined to a small boundary layer we realised that we could look at the problem analytically. This provides us with some generic analytical results on the heating that we can use for any star and any form of heating.

\subsection{Analytic Modelling}

We have investigated the situation and solved the equations analytically using boundary layer approximations to keep only the highest order radial derivatives. We use the same method of Fourier expansion of the variables. We simplify the equations by restricting them to the surface layer and using only the dominant terms in that region. We start by solving the mass conservation equation for incompressible flow

\begin{equation}
\nabla \cdot (\rho \mathbf{u}) =\rm{0}.
\end{equation}
We assume the density variation is negligible and define a circulation variable $\psi$ such that
\begin{equation}
\rho u_{r} = \frac{1}{r^{2}} \frac{\partial \psi}{\partial \mu}
\qquad {\rm{and}} \qquad
\rho u_{\theta} = \frac{1}{r \sqrt{1-\mu^{2}}} \frac{\partial \psi}{\partial r}.
\end{equation}
We now make the variables dimensionless by scaling distance to the star's radius, $R$, and velocities using the viscous diffusion timescale so that $\psi' = \psi R \nu$ becomes dimensionless. Thus the energy equation (5) and vorticity equation (1) become
\begin{equation}
(1-\mu^{2}) \frac{g}{T} \frac{\partial \tilde{T}}{\partial \mu} = \frac{ \nu^{2} }{ R^{3} } \frac{\partial^{4} \psi'}{\partial r^{4}}
\end{equation}
and
\begin{equation}
\frac{\nu R \rho c_{p} N^{2} T}{k g} \frac{\partial \psi'}{\partial \mu} = \nabla^{2} \tilde{T}.
\end{equation}
where $g$ is the acceleration due to gravity, $c_{p}$ the specific heat capacity at constant pressure and $N^{2}$ the buoyancy frequency.

To solve the equations, we expand the variables $\tilde{T}$ and $\psi$ as

\begin{equation}
\tilde{T} \to T_{0}(r) + \mu T_{1}(r) + (2\mu^{2}-1) T_{2}(r)
\end{equation}
and
\begin{equation}
\frac{\partial \psi'}{\partial \mu} = u_{0}(r) + \mu u_{1}(r) + (2\mu^{2}-1) u_{2}(r).
\end{equation}

\begin{figure}[!h]
\plotone{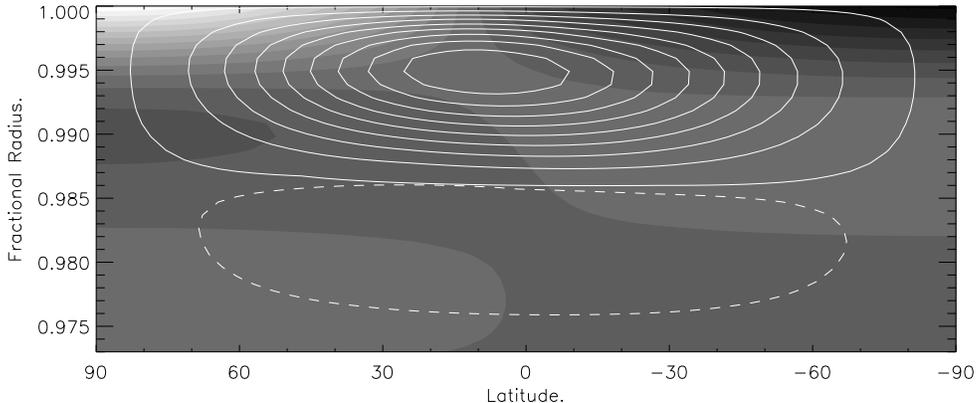}
\caption{Plots of $\tilde{T} (r,\theta)$ and $\psi$ for the solar radiative zone model and the numerical method, the details are identical to those in figure 1. Greyscale levels represent $\tilde{T}(r,\theta)$. With white the highest temperature and black the lowest. Line contours of $\psi$ represent the circulations. Solid lines are clockwise flows, dashed anti-clockwise. The velocities are greatest at the surface and decay with depth as determined by $\delta$. The box is $2.6 \times 10^{11}$ by $6.5 \times 10^{9}$cm. Scaled to real viscosity the y-axis goes from 1.0 to 0.99990.}
\end{figure}

To zeroth order the system reduces to Laplace's equation for the temperature perturbation. For the higher orders we use equations (12)--(16) to obtain equations of the form
\begin{equation}
T_{i}=\delta^{6}_{i}\frac{\partial^{6} T_{i}}{\partial r^{6}},
\end{equation}
when $i > 0$. These have a solution of the form
\begin{equation}
T_{i}={\mathcal{T}}_{i} \exp \Big( \delta_{i} (r-1) \Big) ,
\end{equation}
with similar equations for the flow velocities. The value of $\delta$ is of most interest because it determines the depth of penetration below the surface. With this analysis we obtain

\begin{equation}
\delta_{1} = \Bigg( \frac{2 \rho c_{p} N^{2} R^{4}}{\nu k} \Bigg)^{\frac{1}{6}}
\,\,\, {\rm{and}} \,\,\,
\delta_{2} = \Bigg( \frac{6 \rho c_{p} N^{2} R^{4}}{\nu k} \Bigg)^{\frac{1}{6}}.
\end{equation}
This gives us six solutions for $\delta_{i}$ in the complex plane and we use those that decay with depth. We find the full solution becomes
\begin{equation}
T_{i}(r) ={\mathcal{T}}_{i}  \Bigg( \exp \Big(\delta_{i}(r-1)\Big) + 2 \exp \Big( \frac{1}{2} \delta_{i} (r-1) \Big) \cos \Big( \frac{\sqrt{3}}{2} \delta_{i} (r-1) \Big) \Bigg).
\end{equation}
Then by fitting to the boundary condition (9) we obtain
\begin{equation}
{\mathcal{T}}_{i} = A_{i} \alpha \Big(12 \sigma_{\rm{B}} T^{3} + \frac{ 2 k \delta_{i}}{R}\Big)^{-1}.
\end{equation}

Orders only differ by a numerical factor, for expansion to second order $A_{1}=\frac{1}{2}$ and $A_{2}=\frac{2}{3 \pi}$. The solutions for the flow components $u_{r}$ and $u_{\theta}$ can be deduced from equations (13) and (21). Table 1 shows values of $\delta_{1}$ for various stellar models, computed with the Eggleton code (see Pols et al. 1996). The stars are above $2M_{\sun}$ to ensure fully radiative envelopes. In the numerical study we had to boost the viscosity by a few orders of magnitude to resolve the boundary layer. But in the analytic work we use real values for the viscosity (Spitzer 1962) and so the penetration is shallower, varying as $\delta \propto \nu^{-\frac{1}{6}}$. However if we use the same values as for the numerical calculations the results agree closely. See figure 1 for the comparison of results by this method and those from numerical results.

The temperature perturbations are also listed in table 1, with example flow velocities in table 2. The sources are placed at a distance of $1\,$au from the secondary. This is much larger than the typical separation in an X-ray binary (a few $R_{\sun}$ but we cannot yet deal with such intense irradiation -- see Discussion). For one of the stellar models we show the variation of the values with different illuminating fluxes.

\begin{table}
\caption{Typical values of $\delta_{1}$ and $\mathcal{T}_{\mathnormal{i}}$ from the analytic method. Stellar models were obtained from the Eggleton stellar evolution code. $\delta_{1}$ is dimensionless, $R_{\rm} \delta_{1}^{-1}$ is the penetration depth.}
\begin{center}
\begin{tabular}{|cccccccc|}
\hline
Star Mass & $\delta_{1}$& $R$  & Penetration& $L_{X}$ &$\mathcal{T}_{\rm{0}}$ & $\mathcal{T}_{\rm{1}}$ & $\mathcal{T}_{\rm{2}}$\\
 &  & $/ R_{\sun}$ &  depth  / $R_{\sun}$ & /$\rm{erg}\, \rm{s}^{-1}$ & /mK & /mK & /mK\\
\hline
$2M_{\sun}$ & 2962 & 2.339 & 0.00079 & $10^{33}$ &   0.671  & 0.193& 0.0750\\
 & & & & $10^{35}$ &   67.1 &  19.3& 7.50\\
 & & & & $10^{37}$ &  6710&  1930 & 750 \\
$3M_{\sun}$ & 1965 & 3.001 & 0.00153  &$10^{35}$ &25.1  & 6.59 &  2.54\\
$4M_{\sun}$ & 1866 & 3.417 & 0.00183 &$10^{35}$ & 15.8  &  4.12 &  1.59\\
$6M_{\sun}$ & 1790 & 4.434 & 0.00248 &$10^{35}$ & 6.73 &  1.66 &   0.63\\
$8M_{\sun}$ & 1719 & 5.055 & 0.00294 &$10^{35}$ &  4.55 &   1.11 &  0.43\\
$10M_{\sun}$  & 1657 & 5.769 & 0.00348 &$10^{35}$ & 3.16 &  0.76 &   0.29\\
\hline
\end{tabular}
\end{center}
\caption{Maximum values of velocities, in terms of the sound speed, $c_{\rm{s}}$. Values are from the surface, velocity decays with depth.}
\begin{center}
\begin{tabular}{|cccc|}
\hline
Star Mass & $c_{\rm{s}}$/ $\rm{km \, s^{-1}}$ & $L_{X}$/$\rm{erg}\, \rm{s}^{-1}$  &  $U_{\theta}(\mu=0.3)$/$c_{\rm{s}}$\\
\hline
$2M_{\sun}$ & 12.8 & $10^{33}$ &       $2.99 \times 10^{-2}$\\
 & & $10^{35}$&    2.99\\
 & & $10^{37}$ &   299\\
$3M_{\sun}$ & 16.5 & $10^{35}$ &       1.211\\
$4M_{\sun}$ & 17.6 & $10^{35}$ &       0.780\\
$6M_{\sun}$ & 21.4 & $10^{35}$ &       0.225\\
$8M_{\sun}$ & 23.0 & $10^{35}$ &       0.136\\
$10M_{\sun}$ & 24.3 & $10^{35}$&       0.085\\
\hline
\end{tabular}
\end{center}
\end{table}
\subsection{Discussion}

From the tables it can be seen that a circulation is present even if the irradiating flux is low, although the perturbation is too small to be observed. With larger fluxes a problem arises because the latitudinal flow becomes supersonic. This agrees with the results of Kippenhahn and Thomas but would produce shocks in the flow for which we have not allowed in this work.

The amount of extra flux emitted from the unilluminated side of the star can be estimated by comparison of the strength of the dipole mode to the other terms. We find that about $60\%$ of the flux is emitted from the irradiated surface with $40\%$ from the unilluminated side. This is a smaller difference than found by previous authors and is due to the large zeroth order temperature perturbation. The value is large because we have not allowed the star to expand.

There are two interesting deviations from the work of Kippenhahn and Thomas. First we have much shallower penetration similar to the results of K\i rb\i y\i k and second there is more flux emitted from the opposite side. In our next step we shall test these analytic results for various stars numerically.

In these results it should also be noted that the boundary conditions are simple. While they are a reasonable estimate for illumination by black-body-like stars, if the bulk of the incident radiation is in the form of X-rays then energy will not all be deposited at the surface of the star but at a greater depth. 

\section{Future Work} 

The code calculates 2D perturbations to the hydrostatic equilibrium of any stellar radiative region. However it can only deal with linear perturbations to the thermodynamic variables. Therefore the next phase is to develop the code to model non-linear effects and solve for the full structure, not just perturbations to an assumed hydrostatic equilibrium. In particular in the above study, we have not allowed the star to expand or shrink when heated and we shall need to allow deformation, deviations from spherical shape. We shall also improve upon the boundary conditions to include a model atmosphere as in Tout et al. (1989) and to include external radiation pressure from the irradiating flux, along with the possible deeper penetration of the incident radiation. To study less massive stars we shall also be looking at expanding the code to deal with convection zones by some parameterisation that does not require us to follow the full dynamics of convection.

\end{document}